\documentclass[a4paper]{article}

\usepackage{icrc2013}
\usepackage[english]{babel}
\usepackage{amssymb}

\title{Escape and propagation of UHECR protons and neutrons from GRBs, and the cosmic ray-neutrino connection}

\shorttitle{Escape and propagation of UHECR protons and neutrons from GRBs, and the cosmic ray-neutrino connection}

\authors{
Mauricio Bustamante,
Philipp Baerwald,
Walter Winter
}

\afiliations{
Institut f\"ur Physik und Astrophysik, Universit\"at W\"urzburg, 97074 W\"urzburg, Germany
}

\email{mbustamante@physik.uni-wuerzburg.de}

\abstract{We present a model of ultra-high-energy cosmic ray (UHECR) production in the shock-accelerated fireball of a gamma-ray burst. In addition to the standard UHECR origin from neutron escape and decay into protons, our model considers direct proton emission through leakage from the edges of the accelerated baryon-loaded shells that make up the fireball. Depending on the optical thickness of the shells to photohadronic interactions, the source falls in one of three scenarios: the usual, optically thin source dominated by neutron escape, an optically thick source to neutron escape, or a ``direct escape'' scenario, where the main contribution to UHECRs comes from the leaked protons. The associated neutrino production will be different for each scenario, and we see that the standard ``one neutrino per cosmic ray'' assumption is valid only in the optically thin case, while more than one neutrino per cosmic ray is expected in the optically thick scenario. In addition, the extra direct escape component enhances the high-energy part of the UHECR flux, thus improving the agreement between the predictions and the observed flux.}


\begin{document}
\maketitle

\section{Introduction}

The hypothesis of a common hadronic origin of UHECRs, neutrinos, and photons postulates that photohadronic interactions occur between magnetically-confined UHE protons, shock-accelerated to $\sim 10^{12}$ GeV, and a dense background of photons in the source. The leading contribution is given by the $\Delta^+\left(1232\right)$ resonance, {\it i.e.},
\begin{equation}
 p + \gamma \to \Delta^+\left(1232\right) \to
 \left\{\begin{array}{ll}
  n + \pi^+ \; , &  1/3~\mathrm{of~cases} \\
  p + \pi^0 \; , &  2/3~\mathrm{of~cases} 
 \end{array}\right. ~.
\end{equation}
In the standard picture, the neutrons thus created escape the source and beta-decay as $n \to p + e + \bar{\nu}_e$, with the protons eventually reaching Earth as UHECRs. The charged pions decay and generate UHE neutrinos through
\begin{equation}
 \pi^+ \to \mu^+ + \nu_\mu \; , \;\; \mu^+ \to e^+ + \nu_e + \bar{\nu}_\mu \;,
\end{equation}
while the neutral pions decay as $\pi^0 \to \gamma + \gamma$ to generate the gamma-ray signals observed at Earth. Hence, neutrinos are created in the ratios $\nu_e : \nu_\mu : \nu_\tau = 1 : 2 : 0$, and flavour mixing during propagation transforms this into $1:1:1$ at Earth. We refer to the paradigm of ``one neutrino (of each flavour) per cosmic ray'' as the ``standard case''.

The standard case hinges on two assumptions:
\begin{enumerate}
 \item 
  The protons in the source are perfectly confined by the magnetic field, and only the neutrons may escape.
 \item
  Protons undergo at most one interaction inside the source, while neutrons escape the source without interacting.
\end{enumerate}
Recently, Refs.~\cite{Ahlers:2011jj,Abbasi:2012zw} put the standard case to test and found that GRBs cannot be the sole source of the UHECR protons. However, GRB energetics arguably make them one of the most attractive potential sources of UHECRs and UHE neutrinos. Accordingly, we have explored a generalised emission model in which assumptions (1) and (2) are violated, which we will presently summarise. Further details on our model can be found in Ref.~\cite{Baerwald:2013pu}. 

\section{The GRB fireball model}

In the fireball model, $p\gamma$ interactions occur when relativistically expanding baryon-loaded matter ejecta from a compact emitter collide among themselves, a process in which a fraction of the kinetic energy of the matter is radiated away as protons, photons, and neutrinos. We adopt here a simplified description of the internal collisions in the fireball, following Ref.~\cite{2003LNP...598..393W}. From the measured variability timescale of a GRB, $t_v$ (in the observer's frame), we assume that the central engine emits spherical shells of thickness $\Delta r \simeq c t_v / \left(1+z\right)$ in the source frame, for a burst at redshift $z$. During the first stage of the burst, shells are accelerated by the energy transfer of photons to baryons, eventually reaching a maximum value of the Lorentz factor $\Gamma$, after which the shells coast, each with its own constant maximum velocity. Asssuming fluctuations of the Lorentz factor of the order $\Delta \Gamma / \Gamma \sim 1$, collisions among shells are expected to start at a radius $r_C \simeq 2 \Gamma^2 c t_v / \left(1+z\right)$ from the central engine, signaling the onset of the ``prompt phase'', in which we will focus. The physics of individual collisions are described in detail in Refs.~\cite{Baerwald:2011ee,Hummer:2011ms,Winter:2012xq}.

We compute the secondary ({\it e.g.}, neutrons, neutrinos) injection\footnote{All primed quantities are written in the shock rest frame, SRF. Unprimed quantities refer either to the source or the observer's frame.} $Q^\prime\left(E^\prime\right)$ (in units of GeV$^{-1}$ cm$^{-3}$ s$^{-1}$) coming from photohadronic interactions from the photon and proton densities (GeV$^{-1}$ cm$^{-3}$) as
\begin{equation}
 Q^\prime\left(E^\prime\right) = \int_{E^\prime}^\infty \frac{dE_p^\prime}{E_p^\prime} N_p^\prime\left(E_p^\prime\right) \int_0^\infty c d\varepsilon^\prime N_\gamma^\prime\left(\varepsilon^\prime\right) R\left(x,y\right) \;,
\end{equation}
where $x \equiv E^\prime/E_p^\prime$ is the energy fraction going to the secondary, $y \equiv E_p^\prime \varepsilon^\prime / \left( m_p c^2 \right)$, and $R\left(x,y\right)$ is a ``response function'', which describes the outcome of the interaction as a function of the energies of the incident proton, photon, and secondary, taking into account several $p\gamma$ channels ($\Delta$-resonance, higher resonances, direct pion production) \cite{Hummer:2010vx}.

A broken power law is adopted for the photon density: $N_\gamma^\prime\left(\varepsilon^\prime\right) \propto \left( \varepsilon^\prime / \varepsilon_{\gamma,\mathrm{break}}^\prime \right)^{-\kappa_\gamma}$, with $\kappa_\gamma = \alpha_\gamma \approx 1$ when $\varepsilon_{\gamma,\mathrm{min}}^\prime = 0.2 ~\mathrm{eV} \leq \varepsilon^\prime \leq \varepsilon_{\gamma,\mathrm{break}}^\prime$, $\kappa_\gamma = \beta_\gamma \approx 2$ when $\varepsilon_{\gamma,\mathrm{break}}^\prime \leq \varepsilon^\prime \leq \varepsilon_{\gamma,\mathrm{max}}^\prime = 300 \times \varepsilon_{\gamma,\mathrm{break}}^\prime$, and $N_\gamma^\prime = 0$ otherwise. The break energy $\varepsilon_{\gamma,\mathrm{break}}^\prime = \mathcal{O}\left(\mathrm{keV}\right)$.

For the proton density, Fermi shock acceleration is assumed to generate a non-thermal power-law spectrum of the form $N_p^\prime\left(E_p^\prime\right) \propto \left(E_p^\prime\right)^{-\alpha_p} \times \exp \left[-\left(E_p^\prime/E_{p,\mathrm{max}}^\prime\right)^k\right]$, with $\alpha_p \approx 2$ and $k = 2$. The maximum proton energy $E_{p,\mathrm{max}}^\prime$ is determined by comparing the acceleration timescale $t_\mathrm{acc}^\prime\left(E^\prime\right) = E^\prime / \left(\eta c e B^\prime\right)$ to the dominant loss timescale, {\it i.e.},
\begin{equation}
 t_\mathrm{acc}^\prime\left(E_{p,\mathrm{max}}^\prime\right) = \mathrm{min}\left[ t_\mathrm{dyn}^\prime, t_\mathrm{syn}^\prime\left(E_{p,\mathrm{max}}^\prime\right), t_{p\gamma}^\prime\left(E_{p,\mathrm{max}}^\prime\right) \right] ~,
\end{equation}
where $t_\mathrm{dyn}^\prime \equiv \Delta r^\prime / c$ is the dynamical timescale, $t_\mathrm{syn}^\prime\left(E^\prime\right) = 9 m^4 / \left( 4 c e^4 B^{\prime 2} E^\prime \right)$ is the timescale due to synchrotron losses, $t_{p\gamma}^\prime\left(E^\prime\right)$ is the timescale due to photohadronic interactions, computed numerically as in Ref.~\cite{Hummer:2010vx}, and $\eta$ is the acceleration efficiency.

The photon and proton densities can be determined from the observed radiative flux $F_\gamma$ of the GRB (in units of GeV cm$^{-2}$ s$^{-1}$), with which the isotropic equivalent radiative energy per collision (per shell) can be calculated as $E_\mathrm{iso}^\mathrm{sh} \simeq 4 \pi d_L^2 F_\gamma t_v / \left(1+z\right)$ in the source frame, with $d_L$ the luminosity distance to the source, and $E_\mathrm{iso}^{\prime\mathrm{sh}} = E_\mathrm{iso}^\mathrm{sh} / \Gamma$ in the SRF. Thus, the densities can be normalised through
\begin{equation}
 \int \varepsilon^\prime N_\gamma^\prime\left(\varepsilon^\prime\right) d\varepsilon^\prime = \frac{E_\mathrm{iso}^{\prime \mathrm{sh}}}{V_\mathrm{iso}^\prime} \;, \;\;
 \int E_p^\prime N_p^\prime\left(E_p^\prime\right) dE_p^\prime = \frac{1}{f_e} \frac{E_\mathrm{iso}^{\prime \mathrm{sh}}}{V_\mathrm{iso}^\prime} \;,
\end{equation}
where $V_\mathrm{iso}^\prime = 4 \pi r_C^2 \Delta r^\prime$ is the volume of the interaction region assuming isotropic emission, and $f_e^{-1}$, the ``baryonic loading``, is the ratio between the energy in protons and in electrons. Additionally, the magnetic field can be calculated by assuming that a fraction $\epsilon_e$ and $\epsilon_B$ of energy is carried, respectively, by electrons and by the magnetic field: $B^\prime = \sqrt{ 8 \pi \left( \epsilon_B / \epsilon_e \right) \left( E_\mathrm{iso}^{\prime\mathrm{sh}} / V_\mathrm{iso}^\prime \right) }$. Putting all of this together, the injection spectrum $Q^\prime$ of neutrinos and neutrons can be computed, and the corresponding fluence per shell (in units of GeV$^{-1}$ cm$^{-2}$) is obtained as
\begin{equation}\label{equ:ProtonFluence}
 \mathcal{F}^\mathrm{sh} = t_v V_\mathrm{iso}^\prime \frac{\left(1+z\right)^2}{4 \pi d_L^2} Q^\prime \;, \;\;
 E = \frac{\Gamma}{1+z} E^\prime \;.
\end{equation}
Flavour mixing is implemented for neutrinos, assuming a normal mass hierarchy and the best-fit values of the mixing parameters from Ref.~\cite{Fogli:2012ua}. The total fluence of the burst is obtained by multiplying $\mathcal{F}^\mathrm{sh}$ by the number $N \simeq T_{90} / t_v$ of identical collisions, with the burst duration, $T_{90}$, defined as the time during which $90\%$ of the photon signal is recorded.

\begin{figure}[t!]
 \centering
 \includegraphics[width=0.49\textwidth]{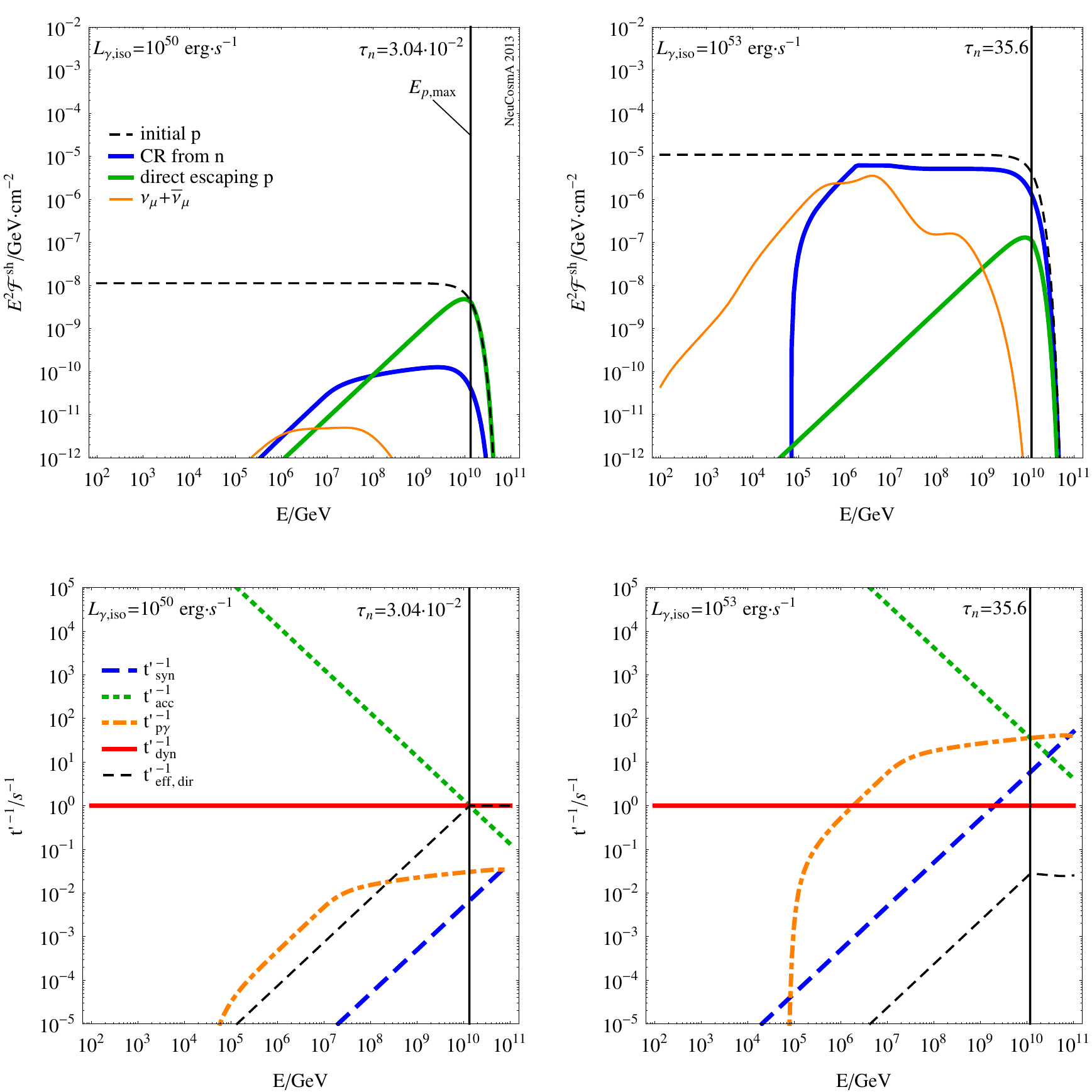}
 \caption{Particle fluences per shell (upper row) and inverse timescales of different processes (lower row) as a function of energy (observer's frame). The left (right) column corresponds to an optically thin (thick) source with $L_{\gamma,\mathrm{iso}} = 10^{50}$ erg s$^{-1}$ ($L_{\gamma,\mathrm{iso}} = 10^{53}$ erg s$^{-1}$). The remaining parameters are set to $\Gamma = 300$, $t_v = 0.01$ s, $\eta = 1$, $\epsilon_e/\epsilon_B = 1$, $f_e = 0.1$, $\alpha_\gamma = 1$, $\beta_\gamma = 2$, $\varepsilon_{\gamma,b}^\prime = 1$ keV, $z = 2$. Adapted from Ref.~\cite{Baerwald:2013pu}.}
 \label{fig:RealburstsCRspectra_singleshell}
\end{figure}

\vspace{-0.35cm}
\section{Optically thin and thick sources, and direct proton escape}

We assume that particles are isotropically distributed within an expanding shell, and that the number of particles that escape the shell is proportional to its volume. Particles from within a shell of thickness $\lambda_\mathrm{mfp}^\prime$ are able to escape without interacting (``mfp'' refers to the mean free path of particles), so that the fraction of escaping particles \cite{Baerwald:2013pu} is $f_\mathrm{esc} \equiv V_\mathrm{direct}^\prime / V_\mathrm{iso}^\prime \simeq \lambda_\mathrm{mfp}^\prime / \Delta r^\prime$.
The mean free paths for protons and neutrons are determined by $\lambda_{p,\mathrm{mfp}}^\prime\left(E^\prime\right) = 
\mathrm{min}\left[ \Delta r^\prime, R_L^\prime\left(E^\prime\right), c t_{p\gamma}^\prime\left(E^\prime\right) \right]$ and $\lambda_{n,\mathrm{mfp}}^\prime\left(E^\prime\right) = \mathrm{min}\left[ \Delta r^\prime, c t_{p\gamma}^\prime\left(E^\prime\right) \right]$, respectively, with $R_L^\prime\left(E^\prime\right) = E^\prime / \left( e B^\prime \right)$ the Larmor radius.
The fluence of directly-escaping protons can then be obtained by multiplying the fluence calculated in equation (\ref{equ:ProtonFluence}) by the fraction $f_\mathrm{esc}$.

\begin{figure*}[t!]
 \centering
 \includegraphics[width=0.85\textwidth]{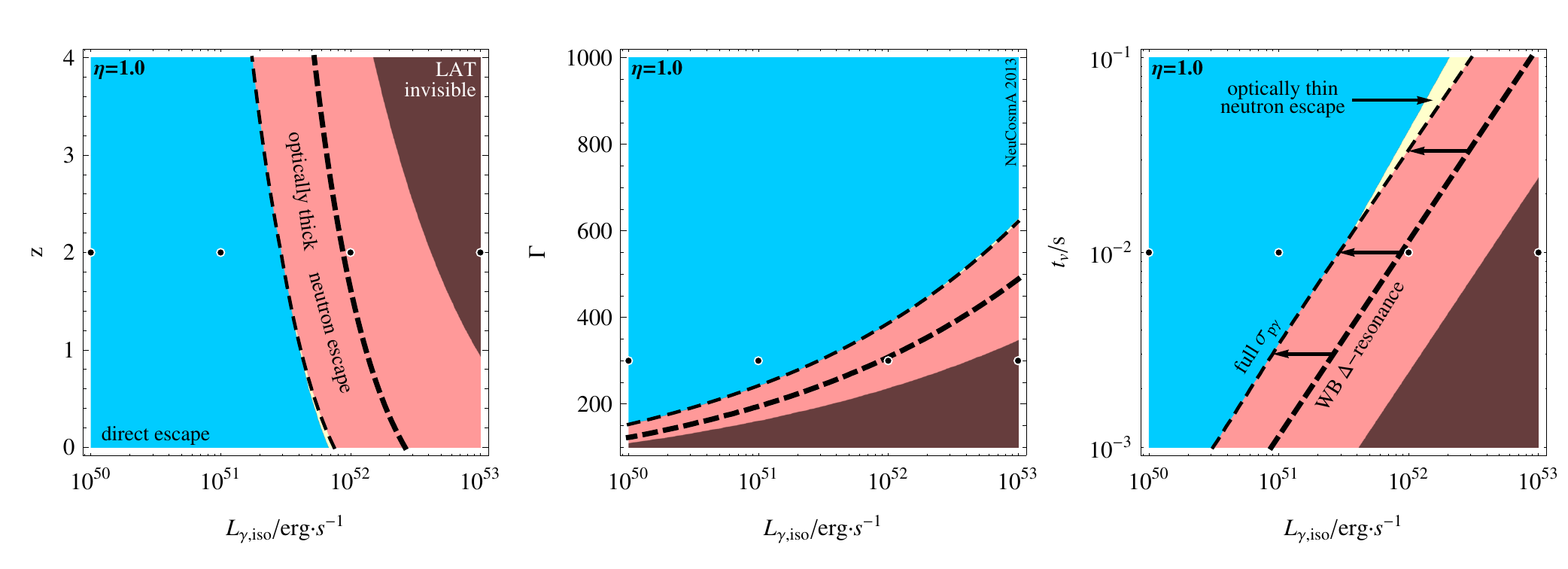}
 \includegraphics[width=0.85\textwidth]{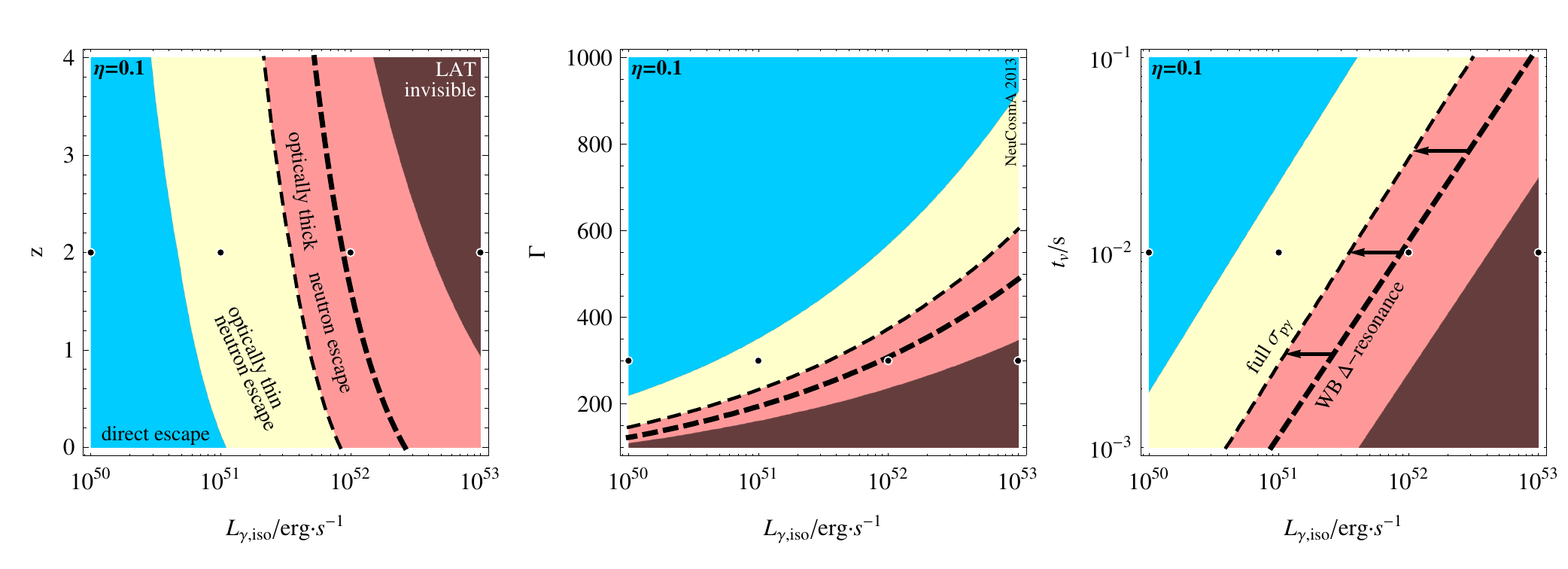}
 \caption{Scan of the GRB parameter space; the unvaried parameters in each plot have the standard values detailed in the caption of figure \ref{fig:RealburstsCRspectra_singleshell}. Differently coloured regions correspond to different emission regimes. The dashed lines mark the interface between the optically thin and optically thick regimes, with the thinner lines resulting from the consideration of the full photohadronic interactions, and the thicker ones, from the $\Delta$-resonance approximation. Within the dark shaded regions labeled ``LAT invisible'', photons created in $\pi^0$ decays will not be able to escape the source due to pair production. The upper and lower rows correspond, respectively, to an acceleration efficiency of $\eta = 1$ and $\eta = 0.1$. Taken from Ref.~\cite{Baerwald:2013pu}.}
 \label{fig:3regimeLisoscan}
\end{figure*}

The source can be characterised by the optical thickness to neutron escape, defined as $\tau_n \equiv \left( t_{p\gamma}^{\prime -1} / t_\mathrm{dyn}^{\prime -1} \right) \vert_{E_{p,\mathrm{max}}}$. If $\tau_n \gtrsim 1$, neutrons (and protons) may interact multiple times in the source and remain confined inside of it ({\it optically thick source}), while the opposite occurs if $\tau_n < 1$ ({\it optically thin source}). Note that, since $t_{p\gamma}^{\prime -1}$ increases with energy, $\tau_n$ will be maximum at $E_{p,\mathrm{max}}$; at lower energies, escape from the source is easier.

Figure \ref{fig:RealburstsCRspectra_singleshell} shows the results for two sample bursts: the left and right columns correspond, respectively, to an optically thin source with $\tau_n = 3.04 \times 10^{-2}$ and an optically thick source with $\tau_n = 35.6$. The upper row shows the particle fluences: ``initial p'' represents the case if all protons were able to directly escape the source over the dynamical timescale, ``CR from n'' represents the protons created from the decay of neutrons that escaped the source, ``direct escaping p'' is the fluence of protons that leaked from the source without interacting, and ``$\nu_\mu + \bar{\nu}_\mu$'' is the muon-neutrino fluence including flavour mixing. Note that to produce this plot only adiabatic losses due to the cosmological expansion have been taken into account during the propagation of both protons and neutrinos. The lower row in figure \ref{fig:RealburstsCRspectra_singleshell} shows the timescales corresponding to the different competing processes at the source. 

For the optically thin source, the maximum proton energy is set by the dynamical timescale, and direct proton escape dominates at $E_{p,\mathrm{max}}$, with most protons being able to escape due to the acceleration efficiency of $\eta = 1$. As a consequence, the associated neutrino fluence is low. For the optically thick source, in comparison, it is the photohadronic timescale which determines the maximum proton energy, and therefore neutron production is enhanced. However, only neutrons lying close to the shell edges are able to escape, which is why the dashed curve cannot be exceeded. On the other hand, neutrinos created everywhere inside the shell are able to free-stream out of it, which leads to a substantially larger neutrino fluence compared to the optically thin case. Three emission regimes are hence identified:
\begin{description}
 \item[Optically thin to neutron escape regime.]
  The standard emission scenario: protons are magnetically confined in the source and photohadronic interactions produce neutrons which are able to escape the source and decay into UHE protons. The charged pion decays lead to the ``one (muon-)neutrino per cosmic ray'' result.
 \vspace{-0.12cm}
 \item[Direct escape regime.]
  Directly escaping protons from the borders dominate the UHECR flux, at least at the highest energies. Neutron production is sub-dominant, and the ``one neutrino per cosmic ray'' paradigm is no longer valid, since more cosmic rays (protons) will be emitted.
 \vspace{-0.12cm}
 \item[Optically thick to neutron escape regime.]
  Neutrons and protons in the bulk of the shell are trapped due to multiple photohadronic interactions, and only those on the borders are able to escape. Neutrino production is enhanced due to the larger number of $p\gamma$ interactions, since neutrinos can escape from anywhere in the shell.
\end{description}

Figure \ref{fig:3regimeLisoscan} shows a numerical scan of the parameter space of GRB emission. For acceleration efficiency of $\eta = 0.1$ (lower row), the three emission regimes are present, with the optically thin regime \---light regions\--- lying close to the standard parameter values: $L_{\gamma,\mathrm{iso}} = 10^{51} \-- 10^{52}$ erg s$^{-1}$, $z \approx 2$, $\Gamma \approx 300$, and $t_v \approx 10^{-2}$ s. Within these regions, the maximum proton energy is lower and, as a result, neutron escape dominates over direct proton escape. Neutrino production is enhanced in the optically thick regime \---light red, or gray, regions\---, where $\tau_n > 1$. In the blue regions, direct proton escape dominates over neutron escape close to $E_{p,\mathrm{max}}$. When perfect acceleration efficiency of $\eta = 1$ is used instead (upper row), the optically thin regions virtually disappear: a higher maximum proton energy allows all protons to directly escape, as long as it is determined by $t_\mathrm{dyn}$. Therefore, for efficient proton acceleration, the ``one neutrino per cosmic ray'' standard result applies only to a very narrow region of parameter space. Additionally, we have marked as ``LAT-invisible'' those regions of parameter space where gamma-rays above $30$ MeV cannot leave the source because they have exceeded the pair-production threshold, and so {\it Fermi}-LAT would not be able to detect these sources.
\vspace{-0.3cm}

\section{UHECR observations}

To explore the effect on the observed UHECR flux of adding the directly-escaping proton component, we have adopted the proton injection spectrum from a sample burst whose parameters make the direct proton escape component dominate at high energies, and the component from neutron decays dominate at lower energies. The transport of protons from the sources to Earth is performed by numerically solving a kinetic equation for the comoving proton density $Y$ (in units of GeV$^{-1}$ cm$^{-3}$) \cite{Ahlers:2009rf,Ahlers:2010fw}:
\begin{equation}
 \dot{Y}
 = \partial_E \left( H E Y \right) 
   + \partial_E \left( b_\mathrm{pair} Y \right)
   + \partial_E \left( b_{p\gamma} Y \right)
   + \mathcal{L}_\mathrm{CR} ~.
\end{equation}
The first term on the rhs takes care of the energy dilution due to the adiabatic cosmological expansion, with $H\left(z\right)$ the Hubble parameter. The second term considers proton energy losses due to pair production on the cosmological microwave (CMB) and infrared/optical photon backgrounds, {\it i.e.}, $p + \gamma \rightarrow p + e^+ + e^-$, while the third term considers losses due to photohadronic interactions. In general, the energy loss rate $b = dE/dt$. The last term, $\mathcal{L}_\mathrm{CR}$, injects protons from the sources at each redshift step, and takes care of the evolution of the source number density with $z$. We have assumed that identical bursts (in the comoving source frame) are distributed in redshift following the GRB rate from Refs.~\cite{Hopkins:2006bw,Kistler:2009mv}.
By evolving the kinetic equation from $z_\mathrm{max} = 6$ down to $z = 0$, the proton flux is obtained as $J\left(E\right) = \left(c/4\pi\right) Y\left(E,z=0\right)$.

\begin{figure}[t]
 \centering
 \includegraphics[width=0.48\textwidth]{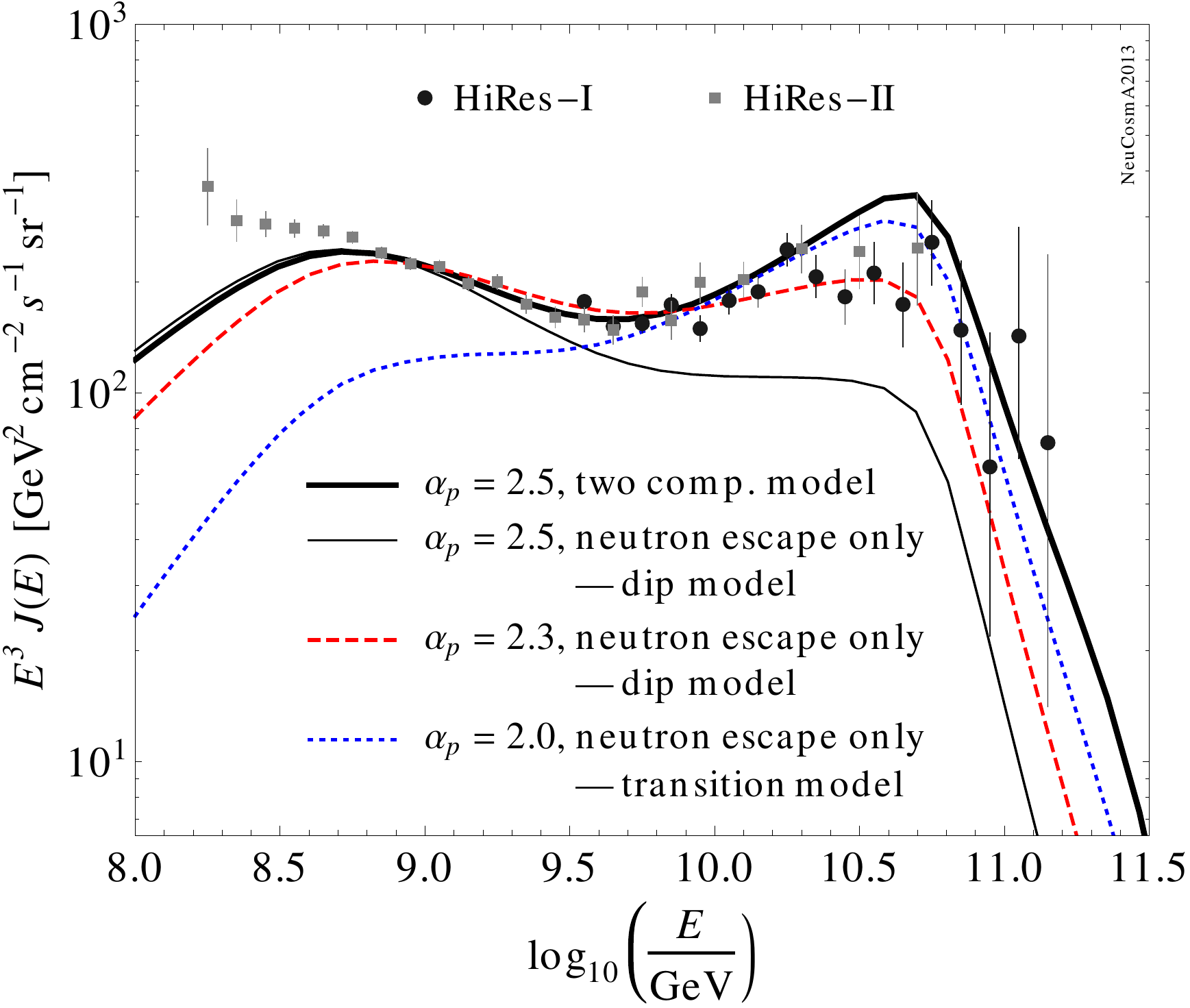}
 \caption{UHECR flux at Earth calculated for different source emission assumptions, normalised to the observed HiRes data \cite{Abbasi:2007sv}. The results for our ``two-component model'' are compared to those calculated using the competing ``dip model'' and ``transition model'', which do not consider direct proton escape. Sources were assumed to follow the GRB redshift evolution \cite{Hopkins:2006bw,Kistler:2009mv} starting at $z_\mathrm{max} = 6$. All bursts were assumed to be identical in the comoving frame, with $\eta = 1$, $t_v = 3.3 \times 10^{-3}$ s, $\Gamma = 10^{2.5}$, $L_{\gamma,\mathrm{iso}} = 7 \times 10^{51}$ erg s$^{-1}$, $\varepsilon_{\gamma,\mathrm{break}}^\prime = 14.76$ keV, $\alpha_\gamma = 1$, $\beta_\gamma = 2$, and $k = 1$, yielding $E_{p,\mathrm{max}} = 1.9 \times 10^{11}$ GeV. Taken from Ref.~\cite{Baerwald:2013pu}.}
 \label{fig:crcomp_GRB}
\end{figure}

Figure \ref{fig:crcomp_GRB} shows the local proton flux resulting from different assumptions of the proton injection spectrum, normalised to the UHECR data points from the HiRes experiment. The ``dip model'' is able to reproduce well the dip in the spectrum due to pair production on the CMB, as well as the ankle, for $\alpha_p \gtrsim 2.5$. However, large values of $\alpha_p$ are difficult to motivate from Fermi shock acceleration. The ``transition model'' can reproduce the ankle using $\alpha_p = 2$, but fails to fit the observations at lower energies, where an extra component, possibly of galactic origin, becomes necessary. Both of these models consider only neutron escape. Finally, our two-component model, shown here for $\alpha_p = 2.5$, is able to fit the observations at lower and higher energies, closely reproducing the spectrum at the dip and ankle. Comparing the curves for $\alpha_p = 2.5$ corresponding to the two-component model and to the dip model, it is clear that the effect of adding the direct proton escape component is to enhance the high-energy part of the spectrum.

\vspace*{-0.3cm}
\section{Summary and conclusions}

We have introduced a model of UHE neutron and proton emission from GRBs in which, depending on the relative dominance of the energy-loss timescales (dynamical, synchrotron, photohadronic), the source can be optically thin or thick to neutron and proton escape, or the emission can be dominated by direct proton escape from the borders of the expanding matter shells. We have tested the validity of the ``one (muon-)neutrino per cosmic ray'' paradigm and found that it is valid only in the optically thin regime, while in the optically thick and direct-escape dominated regimes, either more or fewer neutrinos are created, and this relationship no longer holds. Finally, we have calculated the expected local UHECR flux from our two-component emission model, and found that the addition of the direct-escape component at high energies improves the fit to the experimental data, compared to models with only neutron escape.


\vspace*{0.3cm}
\footnotesize{{\bf Acknowledgments:}{ Work supported by the GRK 1147 ``Theoretical Astrophysics and Particle Physics'', FP7 Invisibles network, Helmholtz Alliance for Astroparticle Physics, and DFG grant WI 2639/4-1. MB acknowledges support from the ICRC organisers.}
\vspace*{-0.4cm}


\end{document}